\documentstyle[12pt]{article}
\begin{document}
\setcounter{secnumdepth}{1}
\title{Remarks on the mean field dynamics of
networks of chaotic elements
\\
}

\author{Kunihiko Kaneko\\
{\small \sl Department of Pure and Applied Sciences,}\\
{\small \sl College of Arts and Sciences,}\\
{\small \sl University of Tokyo,}\\
{\small \sl Komaba, Meguro-ku, Tokyo 153, Japan}\\
}
\date{}

\maketitle

\begin{abstract}

Fluctuations of the mean field of a globally coupled dynamical systems
are discussed.  The origin of hidden coherence is related with
the instability of the fixed point solution of
the self-consistent Perron-Frobenius equation.
Collective dynamics in globally coupled tent maps are re-examined,
both with the help of direct simulation and the Perron-Frobenius equation.
Collective chaos in a single band state, and
bifurcation against initial conditions in a two-band state
are clarified with the return maps of the mean-field, Lyapunov spectra,
and also the newly introduced Lyapunov exponent for the Perron-Frobenius
equation.
Future problems on the collective dynamics are discussed.

\end{abstract}

%\tableofcontents

%\pagebreak

\section{Introduction}

Between the first and last workshops at Como,
we have seen some progress in the study of globally coupled
chaotic systems.  In globally coupled maps,
the studies have elucidated
the clustering\cite{KK-GCM}, the breakdown of the law
of large numbers \cite{KK-MF},
chaotic itinerancy \cite{KK-GCM,KK-CC,CI},
complexity in the partition to clusters \cite{KK-part}, the
existence of high-dimensional tori \cite{Wies,KK-CC,Strogatz}, and
marginal stability with information avalanches\cite{KK-INF}.
The universality of these behaviors
was confirmed by a variety of
mappings \cite{KK-CC,Perez,Sinha},
as well as ordinary differential equations \cite{JJ}.
Related (but partially different) behaviors have been studied in
oscillator-type systems \cite{Wies,KK-CC,Okuda,Strogatz,Kon-KK},
globally coupled Ginzburg-Landau equations \cite{Nakagawa,NK94,Hakim1},
and globally coupled tent maps \cite{KK-MF,Kurtz}
(see also \cite{others}).  Applications to Josephson junction
arrays \cite{Wies,Strogatz,JJ}
and optical systems have also been disussed\cite{opt}, while
relevance of the above
ideas to biological networks has been studied together with extensions to
inhomogeneous systems \cite{Golomb,homeo,KK-rel,Nozawa}.

It is noted that synchronization of globally coupled
oscillators has been studied since the 70's
\cite{Kura-0,Kura-2,Kura-oscil,Daido,Strogatz0}.
However, there is an essential difference between
these earlier studies and recent studies since 1989.  The concern of the
former was synchronization in a system with (randomly) distributed
frequencies (or other parameters) without explicit interest in complex
dynamics (such as chaos), while the main focus of the latter has been
the interplay between chaos and synchronization, where
identical elements can be differentiated and desynchronized due to the
chaotic instability.

The simplest model among such globally coupled
dynamical systems is the globally coupled map (GCM),
originally introduced (\cite{KK-GCM} see also \cite{Wies})
as a mean-field-type extension of the coupled map lattices (CML) \cite{CML}.
The GCM is a discrete-time dynamical system whose elements
interact with all other elements.

Here we focus on the following class of models:
\begin{equation}
x_{n+1}(i)=(1-\epsilon )f(x_n (i))+\frac{\epsilon }{N}\sum_{j=1}^N f(x_n (j)),
\end{equation}
where $n$ is the discrete time and $i$ is the index of elements
($i=$ 1,2, $\cdots ,$ $N$=system size)\cite{KK-GCM}. The
mapping function $f(x)$ is chosen so that it can show chaos.
In the present paper we mainly study
the logistic map
\begin{equation}
f(x)=1-ax^{2},
\end{equation}
or the tent map
\begin{equation}
f(x)=a(0.5-|x-0.5|),
\end{equation}
since there have been investigated in detail as prototypes of
dissipative chaos.

Earlier we reported
the breakdown of the law of large numbers and
the emergence of subtle coherence
\cite{KK-MF} in the desynchronized (turbulent) state.
Here the temporal fluctuations of the mean-field

\begin{equation}
h_n=\frac{1}{N} \sum _i f(x(i))
\end{equation}

are measured with the increase of the system size.  The fluctuations turn
out to decrease only up to some size, beyond which they remain constant.
This is in contrast with the fluctuations in an ensemble of
decorrelated random elements, where the mean square deviation
of the mean-field $h$ should decrease with $N^{-1}$.
The above remnance of the mean field fluctuations, thus, implies
a hidden coherence among the oscillations of the elements.
It should be noted here that this coherence is subtle in nature,
and is neither due to (i) collective behaviors, nor (ii)
the frozen relationships among elements.
In an earlier paper \cite{KK-MF} we have attributed the origin of the breakdown
of the law of large numbers to
the instability
of the fixed point solution of the
Perron-Frobenius (PF) operator.
Here we discuss the relationship in a little more detail
(see also \cite{ershov} and \cite{Kurtz}).

Before going into detailed discussions, we note that the
interaction term in (1) could
be replaced by noise (whose root mean square
is $O(1/\sqrt N)$), if the law of large numbers ( and the central limit
theorem) were valid.  Then, in the limit $N \rightarrow \infty $,
the dynamics of (1) would be reduced to $N$-independent logistic maps
given by $x_{n+1}(i) =(1-\epsilon ) f(x_n (i)) +\epsilon h^*$, where
$h^*=(1/N)\sum _i f(x (i))$ is a constant mean field.

\section{ Instability of the fixed point of the Perron-Frobenius Operator}

In the limit of $N\rightarrow \infty $, the
snapshot distribution $\rho _n (y)$ is defined by the probability
that $x(i)$ assumes the value $y$ ($\rho _n (y) \equiv (1/N) \sum_{j=1}^N
\delta (y-x_n (j))$). This probability distribution can be
calculated with the use of PF operator \cite{Oono,Ruelle}.
It is written as \footnotemark

\footnotetext{There is a typo in eq.4 of \cite{KK-MF}, which should
be read as eq.(5) here.}

\begin{equation}
\rho _{n}(x)=\frac{1}{1-\epsilon}\sum
_{pre=+,-}\frac{\rho_{n-1}(y_{pre})}{|f'(y_{pre})|}
\end{equation}
where the two preimages are given by

\begin{equation}
y_{pre}=f^{-1}(\frac{x-\epsilon \int f(z)\rho _{n-1}(z)dz}{1-\epsilon }),
\end{equation}
and $f^{-1}(y)=\pm \sqrt{\frac{1-y}{a}}$ for
our logistic map.
As was discussed in \cite{KK-MF},
the breakdown of the law of large numbers may be associated with
the instability of the fixed point solution of the
self-consistent Perron-Frobenius (PF) equation.
Indeed the instability can be explained as follows:

If the law of large numbers is valid, the mean-field is constant,
and thus the above distribution is time-independent,
and should be given by the
invariant measure $\rho ^* (y)$ for the logistic map

\begin{equation}
x_{n+1}(i)=(1-\epsilon)f(x_n(i))+\epsilon h^*,
\end{equation}
where $h^*$ is  determined by the self-consistent condition
$h^*=\int f(x)\rho^*(x)dx$.
Since equation (7) is just a one-dimensional map,
the invariant measure is straightforwardly calculated with the
help of the Perron-Frobenius operator for the one-dimensional map.
As is discussed in \cite{KK-MF}, the breakdown of the law of large numbers
implies that the above self-consistent probability solution either
does not exist at all,
or is unstable in case it does exist.   Accordingly we have to
study the existence and stability of the fixed point solution $\rho^*(x)$.

For this study we first consider the one-dimensional map

\begin{equation}
x_{n+1}(i)=(1-\epsilon)f(x_n(i))+\epsilon h_{in},
\end{equation}
where $h_{in}$ is an ``input mean-field" given as a parameter
(which should be determined later self-consistently).
Then we compute the invariant measure $\rho^* _{h_{in}}(x)$
numerically, by using
the Perron-Frobenius operator for the one-dimensional map.
{}From this $\rho^* _{h_{in}}(x)$ we obtain the ``output" mean-field

\begin{equation}
h_{out}=\int f(x)\rho^* _{h_{in}}(x) dx.
\end{equation}

The fixed point solution of our original self-consistent equation
of course should obey $h_{in}=h_{out}$.  The
obtained invariant distribution $\rho^*(x)$ from the one-dimensional map
surely satisfies the self-consistent equation.

Instead of investigating the stability of $\rho^*(x)$ with
regard to the original PF equation,
we check the stability of the self-consistent solution against
a perturbation of the mean field value only.
If this solution is stable, any slight perturbation to $\rho^*(x)$
should decay with time.  Aside from an exceptional form of perturbation,
the value of the mean-field $h$ is altered by it.
Thus, the stability of the fixed point solution
requires $|dh_{out}/dh_{in}|<1$ at $h_{in}=h_{out}$.
In Fig.1, we have plotted $h_{out}$ as a function of $h_{in}$.
As can be seen in Fig.1b), the condition $|dh_{out}/dh_{in}|<1$ is not
satisfied here, which explains the origin of the breakdown of
the law of large numbers.

The complicated behavior in Fig.1 is due to the
existence of windows in the logistic map.  Since windows
exist in any neighborhood of the logistic parameter leading to chaos,
the complicated behavior is expected to be a general feature.
Thus we expect that the law of large numbers is not satisfied
as long as the local mapping has windows.

\vspace{.2in}
---------Fig.1  ----------------
\vspace{.2in}

In the tent map, on the other hand, the relationship between
$h_{out}$ and $h_{in}$ is smooth, and $|dh_{out}/dh_{in}|<1$ at
the unique fixed point $h_{in}=h_{out}$ (see Fig.2).
This confirms the stability of the
fixed point solution of the PF equation for the tent case, while
a rigorous proof needs further studies.
Indeed it has been numerically shown that
the law of large numbers is valid in
this parameter regime of globally coupled tent maps \cite{KK-MF}
( we call this regime completely desynchronized state).

\vspace{.2in}
-------------Fig.2 ----------------------
\vspace{.2in}

Although the Perron-Frobenius approach may be powerful,
the following questions remain to be solved;

\begin{itemize}

\item

How can the crossover size $N_c$ and the remaining correlations be determined
from the PF equation?

Intuitively, the crossover size may be related with the
degree of instability of the fixed point solution.  So far we
have no formulation justifying this intuition,
based on the PF equation.
Numerically it has been found that the variance
$<(\delta h)^2>_{\infty} $ (at the infinity size limit) is inversely
proportional to
$1/N_c$, and that it
is proportional to $\epsilon ^2$ for small coupling $\epsilon$ \cite{KK-MF}.

\item

Dependence on noise

By adding a noise larger than some threshold,
the law of large numbers is known to be recovered.
The threshold seems to be roughly proportional to $<(\delta h)^2>_{\infty}$.
The probability distribution $\rho$ in the stochastic case is written as

\begin{equation}
\rho_{n}(x)=\frac{1}{1-\epsilon}\int dw P(w)
\sum_{pre=+,-}\frac{\rho_{n-1}(y_{pre,w})}{|f'(y_{pre,w})|}
\end{equation}

with

\begin{equation}
y_{pre,w}=f^{-1}(\frac{x-w-\epsilon \int \rho_{n-1}(z)f(z)dz}{1-\epsilon}),
\end{equation}

by using $P(w)$ as the probability of the noise to assume the value $w$.
Indeed $\rho_{n}(x)$ approaches a fixed point distribution
numerically, when the noise is large enough, as is shown in Fig.3.
This agreement is a starting point for further studies on the effect of noise
in terms of the PF equation.

Another mystery remaining to be solved is the anomalous recovery of the
law of large numbers:  When the noise amplitude is not large
enough, the variance of the mean field decays with
$1/N^{\beta}$ with $\beta <1$ \cite{KK-MF}.  This anomalous power may be
due to hierarchical destruction of coherence, which has to
be unveiled in the future.

\item

Another striking mean-field behavior is seen for
a GCM with non-identical, distributed parameters \cite{KK-MF}.  Here
the mean field fluctuation decays with $1/\sqrt{N}$ up to some size,
but then increases to reach the original level of the non-distributed
parameter, with the further increase of size.  So far the origin
is totally unexplained, although it is natural to
expect that the windows existing everywhere in the parameter space
at least play some role.

\end{itemize}

\vspace{.2in}
----------------Fig.3 ----------------------------

\section{Single-band State: High-dimensional Collective Chaos}

In the globally coupled logistic map, transitions among
coherent, ordered (clustered), partially ordered, and turbulent
(desynchronized) phases are
observed, which is rather common in globally coupled dynamical
systems.  In the globally coupled tent map,
the coherent state is stable if $(1-\epsilon )a<1$ is satisfied,
while we have not observed clustered states.

Instead, we have observed desynchronized states
which keep some coherence.
Earlier we reported that the desynchronized states consist of
{\sl single band}, {\sl two-band}, and
completely desynchronized states \cite{KK-MF}.

In a (single) tent map an orbit is always expanding
in contrast with the logistic map.  The motion is linear
in each branch ( here we call $x<0.5$ the left branch and $x>0.5$
the right branch).  Although there are no windows in the chaotic
region, there are band splittings.  For $1<a<\sqrt{2}$, $x_n$ takes a value
larger or smaller than $x^*=a/(1+a)$ (the unstable fixed point),
alternately in time.  Hence two orbits starting from
$x>x^*$ and $x<x^*$ do not mix, and the Perron-Frobenius
equation has a period-2 solution there.

Thus the stability argument of the fixed point
in the previous section does not hold any more, if
the self-consistent tent map
$x_{n+1}=(1-\epsilon ) a(0.5-|x-0.5|)+\epsilon h^* \equiv F(x)$
is in a two-band region (where $h^*$ is
determined self-consistently as in the previous section).
The map $F(x)$ must be in a single-band region, to show a completely
desynchronized state, i.e., $a$ and $\epsilon$ should satisfy
the condition given by $F(F(F(1/2)))>x^*$.
For smaller $a$ or larger $\epsilon$ where the condition is
not satisfied, the map shows two-band splitting,
where the completely desynchronized state disappears
and the globally coupled map yields
the two-band motion as mentioned in \cite{KK-MF}.

In the present section, we study the mean field dynamics
of the single-band state in more detail.
In the single-band state, all elements oscillate almost coherently, that is,
elements always fall within a narrow range (see Fig.18 (b) of \cite{KK-MF}).
Thus the snapshot distribution $\rho_n (x)$ is finite
only within a narrow interval of $x$, and the amplitude of the mean field
dynamics does not decay at all with the increase of the system size.

Here, it is rather straightforward to expect
some collective motion, represented by the mean-field dynamics.
In Fig.4, the return map of the mean field is plotted,
obtained from direct simulation of a GCM with $N=50000$.  We note that
the return map from the PF equation agrees well with Fig.4,
and the scattered points around $x=0.5$ are not due to a finite
size effect.  For $x>0.56$ and $x<0.45$, the plot  exactly
agrees with $f(x)$.  Indeed, if all the elements fall in the same branch
(according to $x>< 0.5$), $\sum_j f(x(j))=f(\sum_j x(j))$,
since $f(x)$ is linear for
each branch.  Thus we get $h_{n+1}=f(h_n)$, when all elements fall within
the same segment of the tent map ( This statement holds generally
for a globally coupled piecewise-linear map).
Since the elements are not in complete synchronization (although close),
they split into two branches when $x_n(i)$ is close to 0.5, leading to the
deviation of the return map from $f(x)$ around 0.5, shown in Fig.4.
Still we should note that the
mean-field  follows the same dynamics as in the coherent case
for most of the time steps $n$ with $x_n$ distant from 0.5.
The ``almost" coherent behavior arises from this use of the same branch.

\vspace{.2in}
-----------Fig.4 ------------------------
\vspace{.2in}

Correlated motion of elements within a small band leads to
collective dynamics.
{}From the time series of the mean field,
we have measured the correlation dimension by increasing the embedding
dimension, expecting to detect a low-dimensional attractor.
So far, however, a saturation in the slope of the correlation integral
\cite{GP} is found neither from the direct simulation of the GCM nor from the
PF equation.  Thus the dynamics of the mean field is still very high.

Lyapunov spectra, on the other hand, have a distinct character.
As is shown in Fig.5, one exponent is much larger than the
other Lyapunov exponents, all of which are positive.
Furthermore, the number of distinct exponents remains one, even if
the number of the elements is increased.

Usually, in spatially extended systems, Lyapunov spectra have a scaling
behavior such
that $\lambda_i =\Lambda(i/N)$ \cite{Manneville,KK-Lyap}.  This is also true
in the desynchronized state of the globally coupled logistic map \cite{KK-MF}.
Recently Nakagawa and Kuramoto have found that in a desynchronized state of the
globally coupled complex Ginzburg Landau equation,
only a few distinctly larger (positive) exponents exist,
whose number remains constant with the increase of the system size.
Other scaled exponents are null or slightly positive,
depending on the parameter \cite{NK94}.
This characteristic feature of the Lyapunov spectra is
similar to our findings for the single-band state here,
although the remaining exponents are always positive in our case.

\vspace{.2in}
--------------Fig.5 ----------------------
\vspace{.2in}

\section{Two-band State}

When the nonlinearity is further increased (or the coupling is decreased)
in the globally coupled tent map, we have a two-band attractor besides
the single-band attractor, as is discussed in \cite{KK-MF}.
Here elements split into two groups, which oscillate out of phases with
each other.
If the oscillation of the elements within the same group
were completely synchronized, this state would be  a two-cluster
attractor as studied in the globally coupled logistic map \cite{KK-GCM}.
Here elements within each group are not synchronized, although they
oscillate within a narrow band, using the same branch.  This
separation is stable: Once the groups are formed, the constituents of
each group are invariant.
Thus, there are many attractors with different partitions,
coded by the number of elements in each group
( for example 45:55, 40:60, 50:50 and so on, for $N=100$) \footnotemark.

\footnotetext{ Note again that clustering in the original sense
is not seen in the globally coupled tent map.}

Now let us study the mean-field dynamics for this 2-band state.
In Fig.6, we have plotted return maps of the mean-field by
changing parameters.  We clearly see collective motions close to
quasiperiodic behavior.

The mean field dynamics changes its nature according to the number of
elements in each group. ``Bifurcations" with
the change of the partition is given in Fig.7, where the return
maps of 5 attractors with different partitions are plotted.
As the imbalance between two groups is larger, the amplitude of
the mean-field dynamics increases, and the motion shows a quasiperiodic
behavior
more clearly.

We have also numerically solved the PF equation starting from a variety
of initial distributions $\rho_0(x)$.  Again, depending on $\rho_0(x)$,
we have obtained a variety of spatial return maps corresponding
to those obtained from the direct simulation (see Fig 8 for
3 examples, although patterns corresponding to other cases are also found).
Note that the width of scattered points remains finite even in the
PF equation ( i.e., the $N \rightarrow \infty$ limit), and thus that the
attractor
is higher than one-dimensional
( i.e., the section of a 2-dimensional torus).

It should be noted that the dynamical state ``bifurcates" with the change of
the number of elements.  This ``bifurcation" is quite analogous to
the cluster bifurcations found in most GCM \cite{KK-GCM}.  Indeed, if
all the elements within the same group were synchronized, our model
would reduce to the two dimensional map

\begin{equation}
X_{n+1}^{\ell}=
(1-\epsilon ) f(X_n ^{\ell})+\epsilon \sum_{j=1}^2(N_j/N) f(X_n ^j).
\end{equation}

Then the bifurcation with $N_j$ would be nothing but that with the change of
the coupling
parameter in the two-dimensional map.  Since elements in the same group
are not synchronized but agree only within a band width $\delta$,
the above reduction is possible only within the precision $\delta$, and
the cluster bifurcation serves only as an approximate explanation of the
change of the return map.

The Lyapunov spectra do not show any remarkable change against
the ``bifurcation" with the number of elements in a group.
All exponents take rather close values, as shown in Fig.5 a) and Fig 9a).
Contrary to the clustering case \cite{KK-GCM}, a
stepwise structure is not observed either, which might be expected if
the oscillation in each group were synchronized ( see Fig. 9).  Thus the above
approximation to two clusters is rather crude, and the deviation in each
group is not negligible for the total dynamics.
We also note that, when the single-band and 2-band states coexist,
the former has a slightly larger maximal Lyapunov exponent (see Fig. 9b), while
the Kolmogorov-Sinai entropy does not show any difference between the two
states,
within our numerical accuracy.

Although collective motion can clearly be seen, we have not yet found
a case where the mean-field dynamics clearly shows a low-dimensional torus.
The slope of the correlation dimension increases beyond 4, and
it is hard to confirm the low-dimensional motion.

\vspace{.2in}

----------Fig.6 ------------------

----------Fig.7 -------------------

--------------Fig.8 -----------------------------------

--------------Fig.9 ------------------------------------

\section{Lyapunov exponent for the Perron-Frobenius equation}

In order to investigate the stability of the mean-field
dynamics, we consider the evolution in a
tangential space of the probability
function $\rho_n (x)$.  First apply
a tiny perturbation $\delta \rho(x)$ to make
$\rho_n'(x)=\rho_n(x)+\delta \rho_n(x)$
($\int \delta \rho (x)=0$ to satisfy the normalization condition)
and see how this $\delta \rho_n(x)$ evolves.
Applying a standard method for calculating the Lyapunov exponent,
we measure the growth rate by\footnotemark

\begin{equation}
lim_{T \rightarrow \infty} lim_{\delta \rho_0 \rightarrow 0} \sum_n^T
\frac{1}{T} log\frac{\int \delta \rho_{n+1}(x)^2 dx}
{\int \delta \rho_n(x)^2 dx}.
\end{equation}

\footnotetext{For computation, $\int \delta \rho_n (x) ^2dx $ is rescaled
to a given tiny value ( say $10~{-5}$) at every time step like in the
usual computation of Lyapunov exponents without the use of the Jacobi
matrix.}

In our self-consistent PF equation, the above exponent is
negative, when $\rho_n (x)$ approaches a fixed point (and
the mean field satisfies the law of large numbers).
On the other hand, the exponent is
positive, when the collective dynamics is chaotic.

The change of the Lyapunov exponents with parameters $a$ and $\epsilon$
is given in Fig. 10.  With the increase of $\epsilon$,
bifurcations from the completely desynchronized state, and
the 2-band state with collective torus, to the (high-dimensional)
collective chaos is seen as the change of the sign of the Lyapunov exponent.
{}From our numerical results by fixing $\epsilon$ at $.3$,
the dependence of the Lyapunov exponent is roughly proportional to
$(a_c-a)$ (for $a<a_c$), with $a_c \approx 1.81$, beyond which the
quasiperiodic motion of mean-field dynamics is seen.
Here the existence of another branch ( for $a<1.6$ ) is
due to the single-band motion.  It is shown that
the single band motion gives a much smaller Lyapunov exponent.
In principle it must be possible to detect some
variation against the partition to groups in the 2-band state, which, so far,
has not clearly been confirmed, as in the case of the Lyapunov spectra from the
GCM in \S 4 \footnotemark.

\footnotetext{
In Fig.10, only the exponents from 5 initial distributions
are measured.  By sampling more initial distributions, we may
expect changes of Lyapunov exponents according to the partition into
two groups in the 2-band state.}

\vspace{.2in}
--------------------Fig.10 ------------------------------
\vspace{.2in}

The Lyapunov spectra of the PF equation
may be a useful tool to analyze the instability of the mean-field.
We have several problems to be studied in future:

\begin{itemize}

\item
Here we have discussed only the maximal exponent, but measuring all
the exponents will be important in order to
detect the high-dimensional collective motions, which
are otherwise hard to confirm.

\item
Possible relationships between the Lyapunov spectra in the original GCM and
the spectra in MF:
The relationship, which one may expect to exist, cannot be so straightforward.
For example, in the completely desynchronized state, all the
exponents in the GCM are positive, but the maximal exponent in the
PF equation is negative.  When the collective motion shows
quasiperiodic motion with a null exponent in the PF,
the exponents in GCM are again positive.
Nontrivial relationships between two Lyapunov spectra should be explored,
for which detailed studies of $N$-dependence of the spectra in GCM may be
necessary.

\item
Since the exponent from PF equation gives the (in)stability
of the mean-field dynamics, one may expect to find some
answers for the remaining questions in \S 2.

\item
Of course, it is interesting to describe the bifurcations
of the mean-field dynamics ( against $a$ and $\epsilon$ as well as
the partition to groups) in terms of the Lyapunov exponents
of the PF equation.

\end{itemize}

\section{Discussion}

In the present paper we have discussed the mean field dynamics of
globally coupled maps.  Some of the earlier reports are discussed
with the use of the Perron-Frobenius equation.
The mean-field dynamics in globally coupled tent maps is
studied in more detail.  The complete desynchronization
in one phase is explained by the stability of the fixed point
of the PF equation, while it shows high-dimensional collective chaos
in a single band state.  In the two-band state the bifurcation from
high-dimensional tori to chaos is observed with respect to
the number of elements in grouping.
Lyapunov analyses from the direct GCM and from the PF equation
are presented.

The globally coupled tent map has some features which distinguish it
from the logistic and other cases,
with regards to the band phases, the mean-field dynamics,
and the complete desynchronization.  Indeed these features
are commonly observed in a GCM with an always expanding piece-wise linear map
(i.e., the absolute value of the slope is always larger than 1).
As an example, we plot the overlaid time series of the GCM (1)
with

\vspace{.1in}

\begin{math}
f(x)=ax (x<0.5); a(x-0.5) (x>0.5),
\end{math}

\vspace{.1in}
in Fig.11.  Again we see single-band, 2-band, (and a higher band),
and completely desynchronized states.  The dynamics belongs to the
same class as the globally coupled tent map.

Quite recently Pikovsky and Kurths reported on the collective
dynamics of globally coupled tent maps with
a "multiplicative coupling" \cite{Kurtz}.
They have presented a beautiful example of clear
quasiperiodic  mean-field dynamics.
Although these behaviors are somewhat similar to
our globally coupled tent map, thus far band-type motion has
not been reported in their example.

Recall that in globally coupled logistic maps
( and in other systems with expansion and contraction),
neither clear mean-field dynamics nor complete desynchronization
is found. Thus we conclude non-trivial violation of the law of large
numbers.
In the globally coupled complex Ginzburg Landau equation,
collective behavior has some features in common with the
single-band state in our globally coupled tent map, although
it has some differences
both from the logistic and tent maps,
as for the number of clusters, and the mean-field dynamics.
Classification of globally coupled dynamical systems waits for further
studies.

Another remaining question is about the topology of couplings.
In locally coupled dynamical systems, collective
quasiperiodic motion has been extensively studied \cite{Chate}.
The collective order of the Ising type has explicitly
been constructed \cite{Sakaguchi},
while the antiferro type order (with a zigzag pattern) is
often found in CML \cite{CML}.  In the former case,
the collective dynamics is  a fixed point, while a periodic or
quasiperiodic collective motion is observed in the latter case.
More generally, such quasiperiodic collective motion coexists
with local chaos in the pattern selection regime of the CML \cite{CML},
as well as in the model of the Benard convection with a large aspect ratio
\cite{Yana}.  The collective order in the two-band state here
can be regarded as the mean-field version of such quasiperiodic
collective order in the zigzag regime of CML.  In a
CML ( with local interaction), a self-consistent Perron Frobenius approach
has been developed \cite{SCPF}.  It will be important to study
the collective dynamics in local CML by using this PF approach.
Last, we note that the clustering in a hypercubic lattice is
discussed, which lies between global and local interactions
\cite{KK-rel}.   Further studies will be necessary for
the connection between locally and globally coupled chaotic systems.

\vspace{.2in}
---------------------Fig.11-------------------------
\vspace{.2in}

{\sl Acknowledgements}

The author would like to thank Y. Takahashi, N. Nakagawa,
T. Konishi, and T. Ikegami for useful discussions.
He would also like to thank F. Willeboordse for critical reading of the
manuscript and illuminating comments.
Thanks are also extended to Dr. S.Ershov for sending me
his unpublished notes.
This work is partially supported by Grant-in-Aids for Scientific
Research from the Ministry of Education, Science, and Culture
of Japan.

%\vspace*{.1in}

\pagebreak

{\em Figure Captions}

\vspace*{.1in}

Fig.1

Relation between $h_{in}$ and $h_{out}$ obtained from the PF
equation for a single logistic map. $a=1.9$ and $\epsilon=0.1$.
For reference the line $h_{in}=h_{out}$ is drawn.

a) for $.1<h_{in}< .5$ with the increment 0.002.
b) Blowup around the region with $h_{in}=h_{out}$,
i.e., $.34<h_{in}<.355$ with the increment 0.00005.

\vspace*{.1in}

Fig.2 Relation between $h_{in}$ and $h_{out}$ obtained from the PF
equation for a single tent map with  $a=1.9$ and $\epsilon=0.1$.
Only the region with $.585<h_{in}<.605$ is plotted,
while the linear dependence (with a small slope) extends
to the interval [0,1].
For reference the line $h_{in}=h_{out}$ is drawn.

\vspace{.1in}

Fig.3  Dynamics of the probability distributions $\rho_n(x)$ for
the PF equation with noise. $a=1.9$, and $\epsilon=0.1$.
The noise is taken from a homogeneous distribution over
$[-\sigma,\sigma]$.  Successive snapshot distributions for
$n=$ 500,501,502,503 are overlaid.
a) $\sigma=0.000625$ b) $\sigma=0.00125$.

\vspace{.1in}

Fig.4
Return Map: $h_{n+1}$ vs $h_{n}$
is plotted over 10000 time steps after transients.  $a=1.5$ and $\epsilon =.3$,
obtained from a direct simulation of globally coupled tent map with $N=50000$.

\vspace{.1in}

Fig. 5
Lyapunov spectra for the globally coupled tent map with $\epsilon=.3$,
and $N=100$, obtained from the average over 5000 steps after transients.
a) $a=1.5$, $a=1.55$, $\cdots$, $a=1.95$, from bottom to top.
b) The first 25 exponents for $N=100$ and 400 ($a=1.5$ ).

\vspace{.1in}

Fig. 6  Return Map $h_{n+1}$ vs $h_{n}$ over 50000 steps (after transients),
obtained  from the
direct simulation of the globally coupled tent map with $N=50000$.
$a=1.5$ and $\epsilon =.3$.

a) $a=1.5$, $\epsilon=.15$ b) $a=1.8$, $\epsilon=.3$
c) $a=1.8$,  $\epsilon=.3$ (by using a different initial condition from b))
d) $a=1.95 $ $\epsilon =.3$, $N=200000$ , only the part with $h_n<0.5$
is plotted, while the other part for $h_n>0.5$ is nothing but a line of
$h_{n+1}=f(h_n)$  as in c).

\vspace{.1in}

Fig. 7 Return Map $h_{n+1}$ vs $h_{n}$ over 50000 steps (after transients),
obtained  from the direct simulation of the globally coupled tent map with
$N=50000$, $a=1.5$ and $\epsilon =.3$.
5 examples from different initial conditions are overlaid.
The number of elements in one group is
50000(E), 38072(D), 34563(C), 28500(B), 25039(A), respectively.

\vspace{.1in}

Fig. 8 Return Map $h_{n+1}$ vs $h_{n}$, over 5000 steps (after transients)
is plotted, obtained from the PF equation, for $a=1.5$ and $\epsilon =.3$.
The PF equation is numerically integrated with the mesh 90000.  3 examples
(A,B,C) from different initial distributions are shown.

\vspace{.1in}

Fig. 9

Lyapunov spectra for the globally coupled tent map with $\epsilon=.3$,
and $N=100$,  obtained from the average over 5000 steps after transients.
 5 examples with different partition to groups are overlaid.

a) $a=1.7$
b) $a=1.8$:
Note the single-band state has a slightly larger maximal exponent.

\vspace{.1in}

Fig.10  The (maximal) Lyapunov exponent for the PF equation,
computed from 90000 meshes, by using the average over
3000 steps.

a)$a=1.5$, while $\epsilon$ is incremented by 0.025.  Results from 3 initial
distributions.  For $\epsilon <.1$, the distribution approaches a
fixed point where the Lyapunov exponent is negative, while
quasiperiodic behavior of the mean-field is observed around $.1<\epsilon <.16$.

b)$\epsilon=0.3$, while $a$ is incremented by 0.025.  Results from 5 initial
distributions. For $a>1.85$, quasiperiodic behavior of the mean-field is
seen.

\vspace{.1in}
Fig. 11 :  Overlaid time series of $x_n(i)$
of the globally coupled piecewise-linear map, with $N=100$,
$a=1.9$ and  $\epsilon =.45$, over the time steps from 10000 to 10050.


\begin{thebibliography}{999}
\bibitem{0}
The present article presents only a part of the talk
presented at Como, while other parts are given in the
succeeding references, as well as \cite{KK-rel} for
the application to biological networks.
After the presentation of the talk, some
related papers (e.g., \cite{Kurtz})
have been published which are also discussed here.
\end{thebibliography}
\end{document}